\documentclass{article}

\usepackage{arxiv}

\usepackage[utf8]{inputenc} 
\usepackage[T1]{fontenc}    
\usepackage{hyperref}       
\usepackage{url}            
\usepackage{booktabs}       
\usepackage{amsfonts}       
\usepackage{nicefrac}       
\usepackage{microtype}      
\usepackage{lipsum}		
\usepackage{graphicx}
\usepackage{natbib}
\usepackage{doi}

\title{PyExperimenter:\\Easily distribute experiments and track results\thanks{This paper is published at the Journal of Open Source Software \citep{tornede2023PyExperimenter}. The final authenticated version is available online at
\url{https://doi.org/10.21105/joss.05149}.}}

\author{ 
    \href{https://orcid.org/0000-0001-9954-462X}{\includegraphics[scale=0.06]{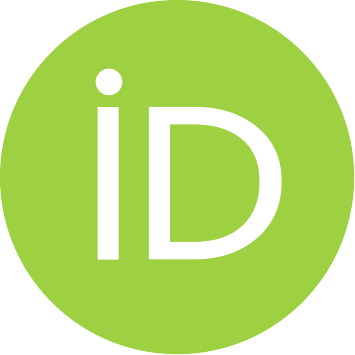}\hspace{1mm}Tanja Tornede},
    \href{https://orcid.org/0000-0002-2415-2186}{\includegraphics[scale=0.06]{orcid.pdf}\hspace{1mm}Alexander Tornede},
    \href{https://orcid.org/0000-0001-8057-4650}{\includegraphics[scale=0.06]{orcid.pdf}\hspace{1mm}Lukas Fehring},
    Lukas Gehring, 
    \href{https://orcid.org/0000-0001-9447-0609}{\includegraphics[scale=0.06]{orcid.pdf}\hspace{1mm}Helena Graf},
    \href{https://orcid.org/0000-0002-1231-4985}{\includegraphics[scale=0.06]{orcid.pdf}\hspace{1mm}Jonas Hanselle}\\
    Department of Computer Science\\
    Paderborn University, Germany\\
    \texttt{\{firstname.lastname\}@upb.de}\\ 
    \texttt{\{fehring2, lgehring\}@mail.uni-paderborn.de}\\
\And
    \href{https://orcid.org/0000-0002-9293-2424}{\includegraphics[scale=0.06]{orcid.pdf}\hspace{1mm}Felix Mohr}\\
    Universidad de La Sabana\\
    Chia, Cundinamarca, Colombia\\
    \texttt{felix.mohr@unisabana.edu.co}\\
\And
    \href{https://orcid.org/0000-0001-9782-6818}{\includegraphics[scale=0.06]{orcid.pdf}\hspace{1mm}Marcel Wever}\\
    MCML, Institut for Informatics\\
    LMU Munich, Germany\\
    \texttt{marcel.wever@ifi.lmu.de}\\
}

\date{April 20, 2023}

\renewcommand{\shorttitle}{\textit{arXiv} Template}

\hypersetup{
pdftitle={PyExperimenter: Easily distribute experiments and track results},
pdfsubject={q-bio.NC, q-bio.QM},
pdfauthor={Tanja Tornede, Alexander Tornede, Lukas Fehring, Lukas Gehring, Helena Graf, Jonas Hanselle, Felix Mohr, Marcel Wever},
pdfkeywords={Experiments, Database},
}

\begin{document}
\maketitle


  
\section{Summary}

\textit{PyExperimenter}\footnote{\textit{PyExperimenter} repository: \url{https://github.com/tornede/py_experimenter}} is a tool to facilitate the setup, documentation, execution, and subsequent evaluation of results from an empirical study of algorithms and in particular is designed to reduce the involved manual effort significantly.
It is intended to be used by researchers in the field of artificial intelligence, but is not limited to those.

The empirical analysis of algorithms is often accompanied by the execution of algorithms for different inputs and variants of the algorithms, specified via parameters, and the measurement of non-functional properties.
Since the individual evaluations are usually independent, the evaluation can be performed in a distributed manner on an HPC system.
However, setting up, documenting, and evaluating the results of such a study is often file-based.
Usually, this requires extensive manual work to create configuration files for the inputs or to read and aggregate measured results from a report file.
In addition, monitoring and restarting individual executions is tedious and time-consuming.

\textit{PyExperimenter} adresses theses challenges by means of a single well defined configuration file and a central database for managing massively parallel evaluations, as well as collecting and aggregating their results.
Thereby, \textit{PyExperimenter} alleviates the aforementioned overhead and allows experiment executions to be defined and monitored with ease.

\begin{figure}[t]
    \centering
    \includegraphics[width=0.9\textwidth]{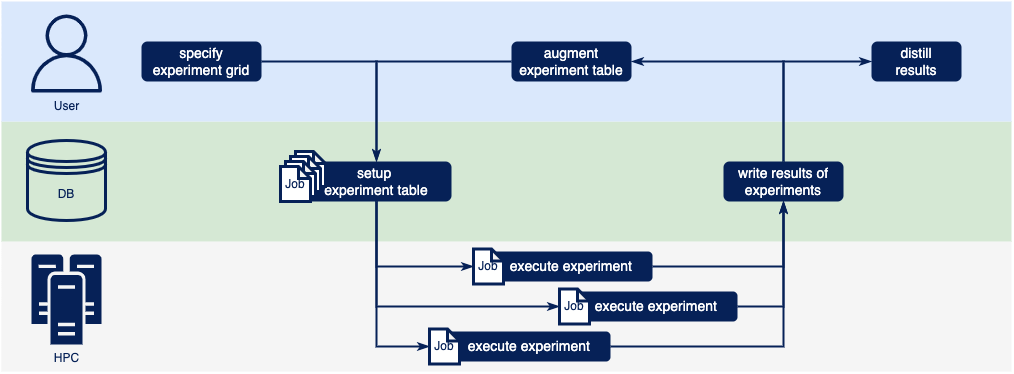}
    \caption{General schema of \textit{PyExperimenter}.}
    \label{fig:usage}
\end{figure}

A general schema of \textit{PyExperimenter} can be found in Figure~\ref{fig:usage}.
\textit{PyExperimenter} is designed based on the assumption that an experiment is uniquely defined by certain inputs, i.e., parameters, and a function computing the results of the experiment based on these parameters.
The set of experiments to be executed can be defined through a configuration file listing the domains of each parameter, or manually through code.
Those parameters define the experiment grid, based on which \textit{PyExperimenter} sets up the table in the database featuring all experiments with their input parameter values and additional information such as the execution status.
Once this table has been created, a \textit{PyExperimenter} instance can be run on any machine, including a distributed system.
Each instance automatically pulls open experiments from the database, executes the function provided by the user with the corresponding parameters defining the experiment and writes back the results computed by the function.
Errors arising during the execution are logged in the database.
In case of failed experiments or if desired otherwise, a subset of the experiments can be reset and restarted easily.
After all experiments are done, results can be jointly exported as a Pandas DataFrame \citep{pandas} for further processing, such as generating a LaTeX table averaging results of randomized computations over different seeds.

\section{Statement of Need}

The recent advances in artificial intelligence have uncovered a need for experiment tracking functionality, leading to the emergence of several tools addressing this issue.
Prominent representatives include Weights and Biases \citep{wandb}, MLFlow \citep{mlflow}, TensorBoard \citep{tensorboard}, neptune.ai \citep{neptune}, Comet.ML \citep{comet}, Aim \citep{aim}, Data Version Control \citep{dvc}, Sacred \citep{sacred}, and Guild.AI \citep{guildai}.
These tools largely assume that users define the configuration of an experiment together with the experiment run itself.
In case of the evaluation of different hyperparameter configurations, this process is suboptimal, since it requires to communicate the hyperparameters through scripts.
This task can become cumbersome to manage as the number of configuration options and desired combinations grows and becomes more complex.
Weights and Biases \citep{wandb}, Polyaxon \citep{polyaxon}, and Comet.ML \citep{comet} allow so-called sweeps, i.e., hyperparameter optimization, albeit in a limited way.
For a sweep, usually hyperparameters that should be optimized are specified along with the desired search domains, and an optimizer can be selected from a pre-defined list to carry out the optimization.
However, the implementation of this functionality usually imposes several restrictions on the way the sweep can be carried out.

In contrast, \textit{PyExperimenter} follows an inverted workflow.
Instead of experiment runners registering experiments to a tracking entity such as a tracking server or database, the experiments are predefined and runners are pulling open experiments from a database.
Similarly, ClearML \citep{clearml} and Polyaxon \citep{polyaxon} support a more generic workflow where experiments are first enqueued in a central orchestration server and agents can then pull tasks from the queue to execute them.
However, both are much more heavyweight than \textit{PyExperimenter} regarding the implementation of both the agents and backend-features. 
Moreover, they are neither completely free nor completely open-source.

In addition to the inverted workflow, a core property of \textit{PyExperimenter} is that the user has direct access to the experiment database, which is usually not the case for alternative tools.
This allows users to view, analyze and modify both the experiment inputs and results directly in the database, although not having to deal with the setup of the database itself.
Sticking to available database technology further does not force the user to learn new query languages just to be able to retrieve files from a database.
Furthermore, \textit{PyExperimenter} offers some convenience functionality like logging errors and the possibility to reset experiments with a specific status such as experiments that failed.

\textit{PyExperimenter} was designed to be used by researchers in the field of artificial intelligence, but is not limited to those.
The general structure of the project allows using \textit{PyExperimenter} for many kinds of experiments as long as they can be defined in terms of input parameters and a correspondingly parameterized function.

\section{Acknowledgements}

This work was partially supported by the German Federal Ministry for Economic Affairs and Climate Action (FLEMING project no.\ 03E16012F) and the German Research Foundation (DFG) within the Collaborative Research Center "On-The-Fly Computing" (SFB 901/3 project no.\ 160364472).

\bibliographystyle{unsrtnat}
\bibliography{references}  

\end{document}